\documentclass[twocolumn,showpacs,preprintnumbers,amsmath,amssymb]{revtex4}

\usepackage{graphicx}
\usepackage{dcolumn}
\usepackage{bm}
\usepackage{mathrsfs}
\usepackage{color}

\begin{document}
\title{Near-threshold correlations of neutrons%
\thanks{Presented at the XXXIII Mazurian Lakes Conference on Physics, Piaski, Poland, September 1-7, 2013}%
}
\author{J. Oko{\l}owicz$^1$, M. P{\l}oszajczak$^2$ and W. Nazarewicz$^{3,4,5}$}

\affiliation{
$^1$ Institute of Nuclear Physics, Radzikowskiego 152, PL-31342 Krak\'ow, Poland\\
$^2$ Grand Acc\'el\'erateur National d'Ions Lourds (GANIL), CEA/DSM - CNRS/IN2P3,
BP 55027, F-14076 Caen Cedex, France\\
$^3$ Department of Physics and Astronomy, University of Tennessee, Knoxville, Tennessee 37996, USA\\
$^4$ Physics Division, Oak Ridge National Laboratory, Oak Ridge, Tennessee 37831, USA\\
$^5$ Institute of Theoretical Physics, University of Warsaw, ul. Ho\.za 69, PL-00-681 Warsaw, Poland}

\date{\today}

\begin{abstract}
\noindent
The appearance  of charged-particle clustering in near-threshold configuration is a phenomenon that can be explained in the Open Quantum System description of the atomic nucleus. In this work we apply the realistic Shell Model Embedded in the Continuum  to elucidate the emergence of neutron correlations in near-threshold many-body states coupled to $\ell=1,2$ neutron decay channels. Spectral consequences of such continuum coupling   are briefly discussed together with  the emergence of complex multi-neutron correlations.
\end{abstract}


\maketitle

\section{Introduction}
It was argued recently that the emergence of charged particle clustering in near-threshold configurations is a genuine consequence of continuum coupling in the Open Quantum System  (OQS) description of nuclear many-body system \cite{cluster}. The coupling of Shell Model (SM) eigenstates via the particle decay channel leads to the formation of the so-called aligned OQS eigenstate, which exhausts  most of the continuum-coupling strength  and is an archetype of a cluster state.  The collectivity of this state is a signature of the instability in an ensemble of all SM states having the same quantum numbers and coupled to the same decay channel. 

What can be said about the nature and appearance of the (multi-)neutron correlations in near-threshold states? The nature of nuclear force rules out the existence of stable like-particle few-body clusters  \cite{tetra-theo}. In heavier nuclei, an interplay between the nuclear mean field and pairing correlations determines the ordering of like-particle decay thresholds and excludes, e.g., the appearance of a 4n-decay channel below 2n-decay channel. Hence, the complex clusters composed of like nucleons can probably be only seen  as  short-lived resonance structures. 

In this paper, we shall study the SMEC wave functions to understand the modification of the spectrum of SM eigenvalues due to the presence of a near-lying  $\ell=1,2$ neutron-decay threshold. The configuration mixing in near-threshold SMEC states will be illustrated by studying exceptional points (EPs) of the complex-extended SMEC Hamiltonian. By investigating the energy- and (complex) interaction-dependence of EPs around the neutron emission threshold, we hope to reveal generic features of the clusterization mechanism for neutrons and its sensitivity to the neutron angular momentum $\ell$.

\section{SMEC: Shell Model for Open Quantum Systems}
To study near-threshold neutron correlations, we employ SMEC that provides a unified description of structure and reactions with up to two nucleons in the scattering continuum \cite{smec}. This configuration interaction approach describes many-body states of an OQS using an effective non-Hermitian Hamiltonian \cite{rf:3,rf:4}. The configuration mixing due to the competition between Hermitian and anti-Hermitian terms is a source of collective features such as, e.g., the resonance trapping \cite{rf:12,rf:13,rf:14}  and super-radiance phenomenon \cite{rf:15,rf:16},  multichannel coupling effects in reaction cross-sectionsm\cite{rf:19}, and charge-particle clustering \cite{cluster}.  

The detailed description of SMEC can be found elsewhere \cite{rf:3,rf:14}. The Hilbert space in this approach is divided into orthogonal subspaces ${\cal Q}_{\mu}$ where index $\mu$ denotes the number of particles  in the scattering continuum. An OQS description of internal dynamics in ${\cal Q}_0$ includes couplings to the environment of decay channels, and is modelled by the energy-dependent  Hamiltonian:
\begin{equation}
{\cal H}(E)=H_0+H_1(E)=H_0+V_0^2h(E),
\label{eq1}
\end{equation}
where  $H_0$ is the Closed Quantum System (CQS) Hamiltonian (here: the SM Hamiltonian), $E$ is the scattering energy, $V_0$ is the continuum-coupling strength,  and $h(E)$ is the coupling term between localized states in ${\cal Q}_0$ and the environment of decay channels. The `external' mixing of any SM eigenstates due to $H_1(E)$ consists of the Hermitian principal value integral describing virtual excitations to the continuum and the anti-Hermitian residuum which represents the irreversible decay out of the internal space ${\cal Q}_0$. 

The SMEC solutions in ${\cal Q}_0$ are found by solving the eigenproblem for the non-Hermitian Hamiltonian (\ref{eq1}). The eigenstates $|\Phi_j\rangle$ of ${\cal H}$ are mixtures of SM eigenstates $|\psi_i\rangle$, and their amplitudes are governed by $V_0$. The continuum-coupling correlation energy of the SM eigenstate $|\psi_i\rangle$,
\begin{equation}
E_{{\rm corr};i}(E) = \langle\psi_i|{\cal H}-H_0|\psi_i\rangle \simeq V_0^2\langle\psi_i|h(E) |\psi_i\rangle, 
\label{eqcorr}
\end{equation}
depends on the structure of the SM eigenstate and the nature of the decay channel. In general, the continuum coupling lowers the binding energies of eigenstates that are close to the decay threshold. As discussed in Refs.~\cite{cluster}, the peak of the continuum coupling correlation energy can be shifted above the threshold due to a combined effect of the centrifugal and Coulomb barriers. 

The information about the configuration mixing is contained in the double poles of the scattering matrix, the so-called exceptional points (EPs) \cite{rf:5}. The mixing of wave functions in CQSs is related to nearby EP(s) of the complex-extended CQS Hamiltonian. Similarly, in OQSs one should analyze the distribution of EPs of the complex-extended OQS Hamiltonian \cite{rf:29}. Since the OQS Hamiltonian (\ref{eq1}) is energy dependent, the complete picture of the near-threshold configuration mixing is contained in exceptional threads (ETs), the trajectories of coalescing eigenvalues $E_{i_1}(E)=E_{i_2}(E)$  of the effective Hamiltonian defined by a complex strength $V_0$ \cite{rf:39}. 

\section{Continuum-coupling correlations near the neutral particle decay channel}
As shown in Ref. \cite{cluster}, the continuum-coupling correlations may lead to collective rearrangements involving many SM eigenstates, i.e., to an instability of certain SM eigenstates near the charge-particle decay threshold. In this section, we shall discuss similar effects, but  for the $J^{\pi}=0^+$ SM eigenstates coupled to the $\ell=1,2$ one-neutron decay channel.  

\subsection{Continuum coupling correlation energy}

\begin{figure}[htb]
\centerline{%
\includegraphics[width=0.45\textwidth]{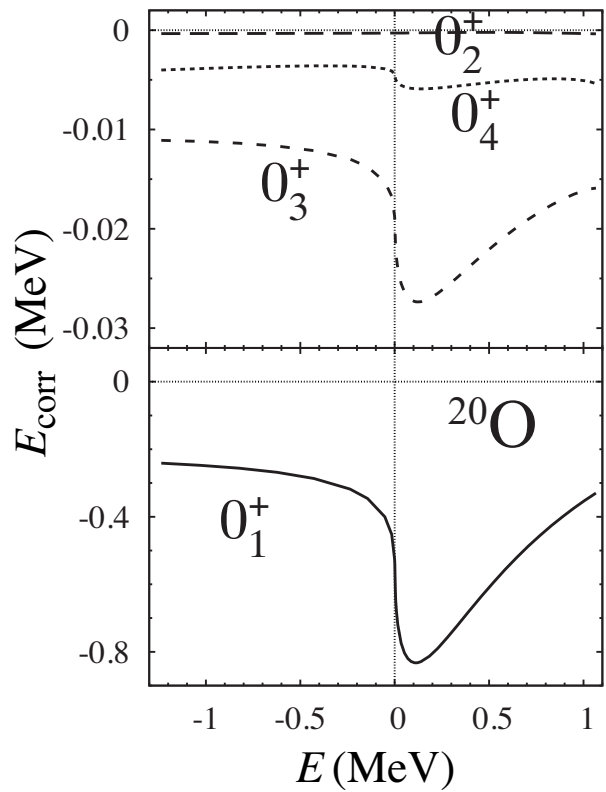}
}
\caption{The continuum-coupling correlation energy $E_{\rm corr}$ (\ref{eqcorr}) to the four lowest  $J^{\pi}=0^+$  shell model eigenvalues in $^{20}$O as  a function of the neutron energy $E$. Bottom: $E_{\rm corr}$ calculated for the  $0_1^+$ ground state (solid line); Top:  The $E_{\rm corr}$ for  $0_2^+$ (long dashed line), $0_3^+$ (short dashed line), and $0_4^+$ (dotted line). The neutron threshold $^{19}$O$(5/2^+)$+n$(\ell = 2)$ at  $E=0$ is indicated by a thin vertical dotted line.}
\label{Fig1}
\end{figure}

Figure \ref{Fig1} shows a typical pattern of the continuum-coupling correlation energy (\ref{eqcorr}) for four SM states of $^{20}$O that are coupled to the $\ell=2$ neutron decay channel.
The SMEC calculations have been carried out  in the  $(0d_{5/2}, 1s_{1/2}, 0d_{3/2})$ model space. For $H_0$, we take the USDB Hamiltonian \cite{rf:br}. The residual coupling between ${\cal Q}_0$ and the embedding one-neutron continuum is generated by the contact force with the continuum coupling strength $V_0=-1000$\,MeV\,fm$^3$. The correlation energy is proportional to $V_0^2$ and  the average density of SM states around the neutron threshold. An interplay between the centrifugal potential and the continuum coupling determines the peak of the continuum-coupling correlation energy, i.e.,  the full width at half maximum $\sigma$ and the centroid $E_{\rm TP}$  of the correlation energy. 

As seen in Fig.~\ref{Fig1}, as compared to the ground-state (g.s.) value, $E_{\rm corr}$ is fairly reduced in excited states. The g.s. curve shows pronounced dependence on the continuum coupling in the vicinity of the threshold. The energy minimum corresponding to maximal correlations is shifted above the threshold by about 0.15 MeV  due to the $\ell=2$ centrifugal barrier. The width of the correlation energy peak is about $\sim$0.5 MeV; this is similar to a typical value of $\sigma$ for the coupling to the $\ell=0$ charged-particle emission threshold \cite{cluster}.

\begin{figure}[htb]
\centerline{%
\includegraphics[width=0.45\textwidth]{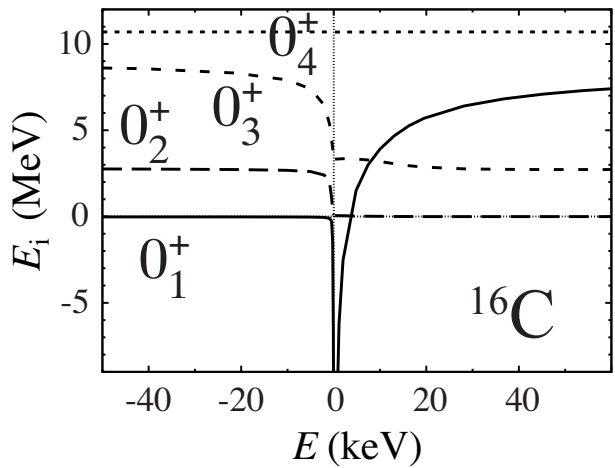}}
\caption{Energies $E_i$ of the four lowest  $J^{\pi}=0^+$  eigenvalues of the effective Hamiltonian (\ref{eq1}) for $^{16}$C  as  a function of the neutron energy $E$. The neutron threshold $^{15}$C$(1/2^-)$+n$(\ell = 1)$ at  $E=0$ is indicated by a thin vertical dotted line. }
\label{Fig2}
\end{figure}

The near-threshold effects of the continuum coupling on SMEC eigenvalues are strongly $\ell$-dependent. Figure \ref{Fig2} shows the dependence of four lowest $0^+$ eigenvalues of SMEC in $^{16}$C on the neutron energy due to the coupling to the neutron decay channel in the partial wave $\ell=1$. 
The SMEC calculations for $^{16}$C have been carried out  in the  $(0p_{1/2}, 0d_{5/2}, 1s_{1/2})$  model space using the ZBM2C Hamiltonian \cite{rf:zu} for $H_0$. The residual continuum coupling is generated by the contact force with  $V_0=-1000$\,MeV\,fm$^3$.  One may notice a successive avoided crossings between SMEC eigenvalues that are indicative of a strong configuration mixing. For the lowest eigenvalue, the dramatic variation of energy around the neutron emission threshold $^{15}$C$(1/2^-)$+n$(\ell = 1)$ is restricted to a range of only few keV. In this narrow energy range, the SMEC eigenvalue has a singular cusp, and the width of the correlation energy peak is less than $\sim1$\,keV. In this energy interval, all excited $0^+$ states are pushed up high in energy with respect to the aligned  $0_1^+$ state. The singular energy behavior in the vicinity of the threshold is not visible for higher eigenvalues. 

\subsection{Configuration mixing near the neutral particle emission threshold}
A comprehensive picture of the near-threshold  mixing of SMEC eigenstates can be obtained by investigating properties of the ETs of the complex-extended SMEC Hamiltonian \cite{cluster}. 
\begin{figure}[htb]
\centerline{%
\includegraphics[width=0.45\textwidth]{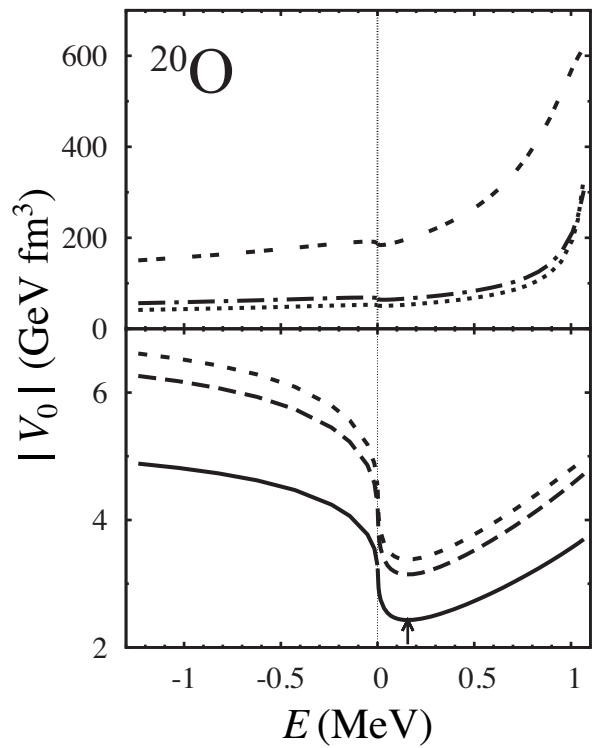}}
\caption{Exceptional threads $|V_0|(E)$ for the $0^+$ eigenvalues of the complex-extended SMEC Hamiltonian (\ref{eq1}) in $^{20}$O. Only exceptional threads corresponding to decaying resonances are shown. Each point along an exceptional thread is an exceptional point  labelled by the neutron energy with respect to the 
$^{19}$O$(5/2^+)$+n$(\ell = 2)$ threshold. There are six threads altogether. Three of them (bottom panel), involving the aligned  $0_1^+$ state, are found in the physical range of $|V_0|$ values and correspond to a coalescence of $0_1^+-0_2^+$ (solid line),  $0_1^+-0_3^+$ (long-dashed line), and $0_1^+-0_4^+$ (short-dashed line) eigenvalues. The remaining three threads (top panel), $0_2^+-0_3^+$,  $0_2^+-0_4^+$, and $0_3^+-0_4^+$, correspond to very large values of $|V_0|$ and have no influence on the eigenstate mixing. The arrow indicates the turning point of the exceptional thread corresponding to the maximal continuum coupling.}
\label{Fig3}
\end{figure}
\begin{figure}[htb]
\centerline{%
\includegraphics[width=0.45\textwidth]{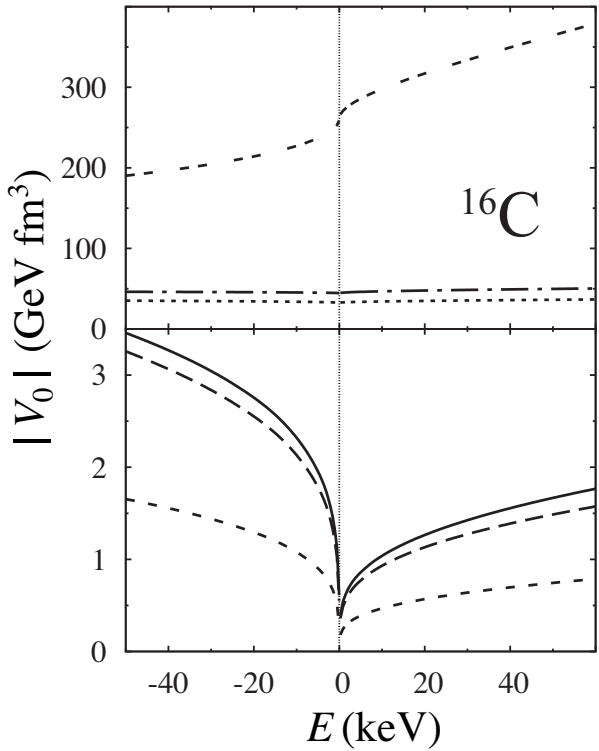}}
\caption{Similar as in Fig. \ref{Fig3} but for $^{16}$C shell model eigenstates coupled to the neutron decay channel $^{15}$C$(1/2^-)$+n$(\ell = 1)$.}
\label{Fig4}
\end{figure}
The energy dependence of ETs in a vicinity of the threshold is shown in Fig. \ref{Fig3} for the eigenstates of $^{20}$O coupled to the  $^{19}$O$(5/2^+)$+n$(\ell = 2)$ channel, and in Fig. \ref{Fig4} for the eigenstates of $^{16}$C coupled to the $^{15}$C$(1/2^-)$+n$(\ell = 1)$ channel.
We consider four $0^+$ SM eigenstates; hence, six different ETs for decaying resonances. All ETs that are relevant for the collective mixing of SM states involve the aligned eigenstate $0_1^+$. They exhibit a minimum of $|V_0|$ (the turning point, TP) at the same neutron energy.  At this energy, the collective mixing of SM states is maximal. The remaining ETs are found at exceedingly large values of $|V_0|$ (see the upper panels of Figs. \ref{Fig3} and \ref{Fig4}) and have no influence on the configuration mixing. In the case of $^{20}$O, the TP  energy $E_{\rm TP}\simeq0.18$\,MeV lies above the $\ell=2$ neutron emission threshold. The dependence of $|V_0|(E)$ around the turning point for those ETs that involve the  $0_1^+$ eigenstate is asymmetric, and significant configuration mixing in SMEC wave functions is predicted up to $E\simeq1.2$\,MeV. 

In the case of $^{16}$C, the turning point is seen at the $\ell=1$ neutron emission threshold for all ETs containing the aligned  $0^+$ eigenstate. The energy dependence in the vicinity of the threshold of these ETs exhibits a cusp-like behavior. 

\begin{figure}[htb]
\centerline{%
\includegraphics[width=0.45\textwidth]{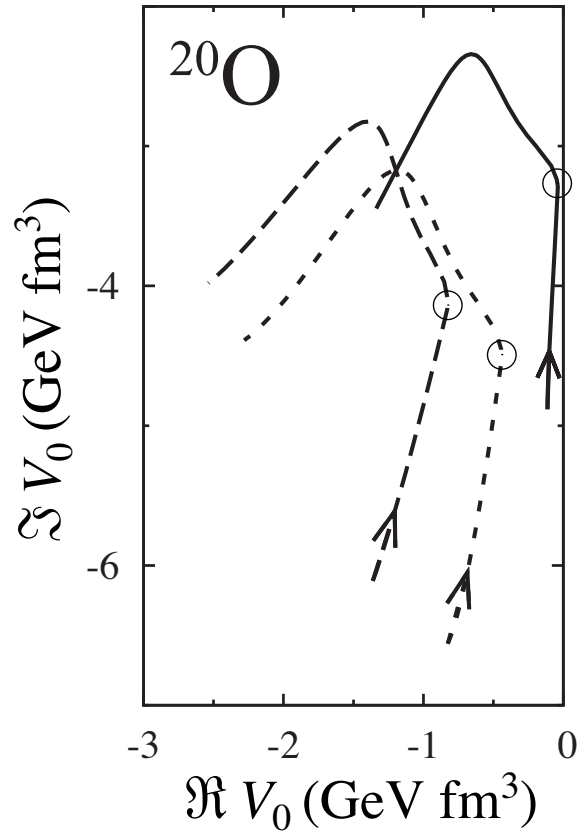}}
\caption{Exceptional threads  in the complex-$V_0$ plane for $0^+$ eigenfunctions of the SMEC Hamiltonian in $^{20}$O coupled to the  $\ell=2$ neutron decay channel $^{19}$O$(5/2^+)$+n$(\ell = 2)$. Each point at the exceptional thread corresponds to an exceptional point at a definite energy $E$. The  threshold position ($E=0$) is indicated with a circle for each thread. Only the exceptional threads formed by a coalescence of ($0_1^+$-$0_2^+$) (solid line), ($0_1^+$-$0_3^+$) (long-dashed line), and ($0_1^+$-$0_4^+$) (dashed line) eigenvalues of decaying resonances at low values of $|V_0|$ are shown. Arrows indicate the direction of increased  neutron energy.}
\label{Fig5}
\end{figure}

\begin{figure}[htb]
\centerline{%
\includegraphics[width=0.45\textwidth]{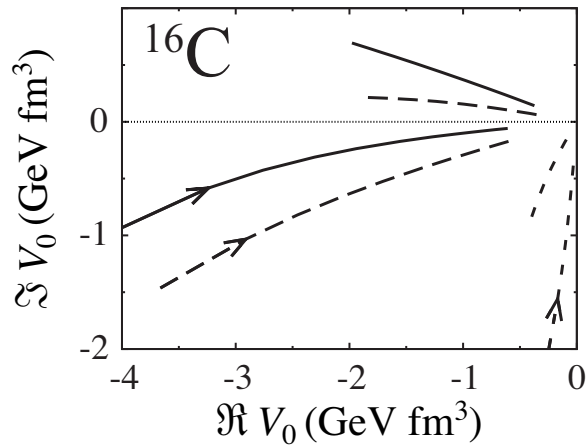}}
\caption{Similar as in Fig. \ref{Fig5} but for $^{16}$C SMEC eigenfunctions coupled to the neutron decay channel $^{15}$C$(1/2^-)$+n$(\ell = 1)$.}
\label{Fig6}
\end{figure}

Another view on the configuration mixing around the neutron decay threshold is provided in Figs. \ref{Fig5} and \ref{Fig6}, which show the ETs in the complex-$V_0$ plane for the  $0^+$ states coupled to $\ell=2$ and $\ell=1$ neutron decay channels, respectively. Each ET in these figures involves the aligned $0_1^+$ state, and each point on trajectories in the complex-$V_0$ plane corresponds to a different neutron energy. 

Figure~\ref{Fig5} shows that the pattern of most important ETs involving the aligned $0_1^+$ state  has two cardinal points: the threshold (denoted by a circle)  and the turning point, which is the point of the closest approach of the ET to the origin of the complex-$V_0$ plane. At the  $E=0$ threshold, the real part of $V_0$ is closest to the axis ${\cal R}eV_0=0$ for each ET. For $0<E<E_{\rm TP}$, the real part of $V_0$ begins to move away from the  ${\cal R}eV_0=0$ axis, whereas the imaginary part of $V_0$ continues to approach rapidly the real-$V_0$ axis. As a result, the strongest collectivization of SM eigenstate and, consequently, the largest continuum-coupling energy correction for $\ell=2$, takes place at the TP, and not at the neutron emission threshold. For energies above the turning point, both $|{\cal R}eV_0|$ and $|{\cal I}mV_0|$ increase monotonously and the collectivization of SM eigenstates  gradually vanishes. 

The radical change is seen when decreasing the neutron angular momentum  from $\ell=2$ to $\ell=1$ (Fig. \ref{Fig6}). In this case, the whole segment of ETs (between $E=0$ and $E=E_{\rm TP}$) disappears, and the TP coincides with the neutron emission threshold. One can see that for energies below the threshold, all ETs, which involve the aligned state approach  $V_0=0$; this is another signature of the extreme collectivization of the aligned SMEC eigenstate. Above the threshold, both $|{\cal R}eV_0|$ and $|{\cal I}mV_0|$ increase rapidly with  neutron energy, and the coupling of SM eigenstates through the decay channel disappears.

We assert that the collective mixing of SM eigenstates via the aligned state is the reason for strong multi-neutron correlations in near-threshold states. 
For neutral-cluster configurations, the observation of clustering depends on whether (i) a SM  state lies within the narrow  energy window resulting from  the competition between the continuum coupling and the centrifugal potential; and (ii)  its lifetime is sufficiently long to be detected. For SM states coupled to the cluster decay channel in low partial waves ($\ell=0, 1$), the window of opportunity for neutral-particle clustering is situated in a narrow energy range around the turning point at the neutral-cluster decay threshold. For higher partial waves ($\ell\geq 2$), this energy window moves above the neutral-cluster decay threshold, but an experimental observation of  clustering is compromised by the rapid decrease of lifetimes of such configurations.  

\section{Conclusions}
The existence of multi-neutron clustering is an old problem in low-energy nuclear physics.  Recently, this problem has been revived in Ref.~\cite{tetra}, who reported the observation of neutron clusters in the disintegration of beryllium and lithium nuclei. It was claimed that correlated four neutrons observed in the disintegration of $^{14}$Be could be related to a four-neutron halo in this nucleus. This report spurred significant theoretical work, which rejected the existence of bound tetraneutron as being incompatible with our understanding of nuclear forces \cite{tetra-theo} (see, however, a discussion in Ref. \cite{berzel} of a two-dineutron molecule). 
The isospin structure of the nuclear force prevents the existence of stable nuclei made of like-nucleons. It also explains the energetic order of decay channels, e.g., ensures that a 4n-emission threshold is above 2n- and/or 1n-emission thresholds. However, the nature of nuclear forces does not preclude the manifestation of tetraneutron correlations in the vicinity of the 4n-emission threshold. The tetraneutron clustering may arise as a consequence of the collective coupling of many-body resonances via the 4n-emission channel. Since at energies close to the 4n-emission channel both 1n- and 2n-emission channels are open, the tetraneutron correlations can be influenced by the coupling to these channels as well. 

The mechanism responsible for the creation of multi-neutron cluster states is analogous to that behind the presence of charged clusters and involves the aligned near-threshold state.  The degree of collectivity of the aligned state depends on the strength of the continuum coupling, the density of SM states, the structure of the emission channel, and the energy of the aligned state relative to the threshold. For  $\ell=1$ neutron decay channels, the collectivity of the aligned state is directly related to the energy gap between the aligned state and  higher-lying states with the same quantum numbers. For channels with higher angular momenta ($\ell\geq2$), the optimal energy window for the formation of a collective multi-neutron state is pushed above the decay threshold, and the possibility to find such a state is small. 

We have demonstrated that the point of the maximum continuum-coupling energy is related to the turning point of all ETs involving the aligned state. This point coalesces with the emission threshold for SM states coupled to $\ell=1$ neutron decay channels. 
The energy of the turning point  weakly depends on the continuum-coupling strength and the masses of  particles involved. Hence, contrary to the charged-particle clustering predicted to disappear in heavier nuclei \cite{cluster}, the multi-neutron correlations around the $\ell=1$ multi-neutron decay threshold are expected to  appear along the whole  nuclear landscape. Unfortunately, the presence of  strong multi-neutron correlations is expected only in  a narrow energy range  around the threshold thus making their experimental detection challenging. The closest approximation to a genuine multi-neutron cluster states are 2n-halos. However, 2n-separation energies in all experimentally studied cases are too large to speak of genuine dineutron cluster configurations \cite{papa}. The experimental observation of  multi-neutron clustering remains, therefore,  a challenging open question for future investigations.

\vskip 0.4truecm
\noindent
{\bf  Acknowledgements}\\
This work was supported in part by the MNiSW grant No. N N202 033837, the Collaboration COPIN-GANIL on physics of exotic nuclei, the Project SARFEN (Structure And Reactions For Exotic Nuclei) within the framework of the ERANET NuPNET, FUSTIPEN (French-U.S. Theory Institute for Physics with Exotic Nuclei) under DOE grant number DE-FG02-10ER41700, and by the DOE grant DE-FG02-96ER40963 with the University of Tennessee.


\begin{thebibliography}{}
\bibitem{cluster}
J. Oko{\l}owicz, M. P{\l}oszajczak and W. Nazarewicz, Prog. Theor. Phys. Supplement {\bf 196} (2012) 230;\\
J. Oko{\l}owicz, W. Nazarewicz and M. P{\l}oszajczak, Fortschr. Phys. {\bf 61} (2013) 66.
\bibitem{tetra-theo} S.C. Pieper, Phys. Rev. Lett. \textbf{90} (2003) 252501.
\bibitem{smec} K. Bennaceur, F. Nowacki, J. Oko{\l}owicz and M. P{\l}oszajczak, Nucl.\ Phys.\ A \textbf{651} (1999) 289;\\
J. Rotureau, J. Oko{\l}owicz and M. P{\l}oszajczak, Nucl.\ Phys.\   A \textbf{767} (2006) 13.
\bibitem{rf:3} 
J. Oko{\l}owicz, M. P{\l}oszajczak and I. Rotter, Phys.\ Rep.\ \textbf{374} (2003) 271.
\bibitem{rf:4} 
A. Volya and V. Zelevinsky, Phys.\ Rev.\ C\textbf{74} (2006) 064314.            
\bibitem{rf:12} P. Kleinw\H{a}chter and I. Rotter, Phys.\ Rev.\ C \textbf{32} (1985) 1742;\\
I. Rotter, Rep.\ Prog.\ Phys.\  \textbf{54} (1991) 635.
\bibitem{rf:13} V.V. Sokolov and V.G. Zelevinsky, Phys.\ Lett.\ B \textbf{202} (1988) 10;\\
V.V. Sokolov and V.G. Zelevinsky, Nucl.\ Phys.\ A \textbf{504} (1989) 562.
\bibitem{rf:14} S.~Dro\.z\.d\.z, J. Oko{\l}owicz, M. P{\l}oszajczak and I. Rotter, Phys.\ Rev.\  C \textbf{62} (2000) 24313.
\bibitem{rf:15} R.H. Dicke, Phys.\ Rev.\ \textbf{93} (1954) 99.
\bibitem{rf:16} N. Auerbach and V.G. Zelevinsky, Rep.\ Prog.\ Phys.\ \textbf{74} (2011) 106301.
\bibitem{rf:19} A.I. Baz, Soviet\ Phys.-JETP\ \textbf{6} (1957) 709;\\
R.G. Newton, Phys.\ Rev.\ \textbf{114} (1959) 1611;\\
C. Hategan, Ann.\ Phys.\ (NY)\ \textbf{116} (1978) 77.
\bibitem{rf:5}  T. Kato, {\em Perturbation Theory for Linear Operators} (Springer Verlag, Berlin, 1995);\\
M.R. Zirnbauer, J.J.M. Verbaarschot and H.A. Weidenm\"{u}ller, Nucl.\ Phys.\  \textbf{A411} (1983) 161;\\
W.D. Heiss and W.-H. Steeb, J.\ Math.\ Phys.\ \textbf{32} (1991) 3003.
\bibitem{rf:29} J. Oko{\l}owicz and M. P{\l}oszajczak, Phys.\ Rev.\ C\ \textbf{80} (2009) 034619.
\bibitem{rf:39} J. Oko{\l}owicz and M. P{\l}oszajczak, Acta\ Physica\ Polonica\ B \textbf{42} (2011) 451.
\bibitem{rf:br} B.A. Brown and W.A. Richter, Phys.\ Rev.\ C \textbf{74} (2006) 034315.
\bibitem{rf:zu} A.P. Zuker, B. Buck and J.B. McGrory, Phys.\ Rev.\ Lett.\ \textbf{21} (1968) 39.
\bibitem{tetra} F.M. Marqu{\'e}s et al., Phys. Rev. C \textbf{65} (2002) 044006.
\bibitem{berzel} C.A. Bertulani and V.G. Zelevinsky, J. of Physics G \textbf{29} (2003) 2431.
\bibitem{papa} G. Papadimitriou, A.T. Kruppa, N. Michel, W. Nazarewicz, M. P{\l}oszajczak and J.
Rotureau, Phys. Rev. C \textbf{84} (2011) 051304(R).



\end{thebibliography}
\end{document}